\magnification1200
\parindent0pt
\baselineskip=1.4\baselineskip
\input pictex
\input amssym.tex

\def\dzien{\number\day\space%
\ifcase\month \or
 stycznia \or lutego \or marca \or kwietnia \or maja \or czerwca \or
 lipca \or sierpnia \or wrzeœnia \or paŸdziernika \or listopada \or
 grudnia \fi
 \number\year}
\newcount\t
\newcount\h
\newcount\m
\def\godzina{\t=\number\time \h=\t
\divide\h by60 \the\h:%
\multiply\h by60
\advance\t by-\h \the\t}

\def\plik{Sieradzki$\_$Zieli\'nski}
\footline{\hss{\sl \plik;\ \folio}}

\def\frac#1#2{{#1\over#2}}
\def\binom#1#2{{#1\choose#2}}

\centerline{\bf An example of application of optimal sample allocation}
\centerline{\bf in a finite population}

\vskip1truecm
\centerline{\bf Dominik Sieradzki}
\centerline{\bf Wojciech Zieli\'nski}
\bigskip
\centerline{Department of Econometrics and Statistics}
\centerline{Warsaw University of Life Sciences}
\centerline{Nowoursynowska 159, PL-02-787 Warszawa}
\centerline{e-mail: dominik$\_$sieradzki@sggw.pl}
\centerline{e-mail: wojciech$\_$zielinski@sggw.pl}
\centerline{http: wojtek.zielinski.statystyka.info}

\vskip1.5truecm

{\bf Summary.} The problem of estimating a proportion of objects with particular attribute in a finite population is considered. This paper shows an example of the application of estimation fraction using new proposed sample allocation in a population divided into two strata. Variance of estimator of proportion which uses proposed sample allocation is compared to variance of the standard one.

\bigskip

{\bf Keywords}: survey sampling, sample allocation, stratification, estimation, proportion

{\bf JEL Classification:} C83, C99

\bigskip

{\bf Introduction.} In microeconomics, the main subject of interest is human as a managing individual, whereas macroeconomics places the greatest emphasis on households and enterprises (Bartkowiak, 2003). Such objects frequently form multi-million populations. Due to amount of costs it is impossible to subject the population of interest to exhaustive sampling, even for Statistical Office. In economics populations consist of a finite number of units. Survey sampling deals with finite populations. Therefore a sample is drawn from the population. When sampling, two types of errors can be distinguished: sampling error and non-sampling error. Non-sampling error is associated with the non-response problem. Proposal on how to deal with such an issue can be found in Hansen and Hurwitz (1946) or Chaudhuri et al. (2009). This article is focused on sampling error, hence it is assumed that responses were obtained from all of the chosen units in the sample. The sampling error, among others, depends on sampling scheme. In the next part of this paper an example of application of sample allocation proposed in Sieradzki and Zieli\'nski (2018) is presented.
In economics the aim of the research is often to inference about dychotomus occurences, for example support for a particular party or candidate in elections (Szreder, 2010), unemployment rate (Hada\'s-Dyduch, 2015), farmers' decision about production credit and EU measures (using these funds or not) (Roszkowska-M¹dra and Ma\'nkowski, 2010) or deciding on ecological farming (Sieradzki and Stefa\'nczyk, 2017). Consider a problem of support for a particular candidate in the elections. Main issue to consider in the study is a population ${\cal U}=\left\{u_1,\dots,u_N\right\}$ which contains a finite number of $N$ people who may vote. In this population a number of people who support a particular candidate is observed. All the units in this population could be considered as a vector $Y=(Y_1,\ldots,Y_N)$, where $Y_k=1$ if $k$-th person supports a candidate and $Y_k=0$ if $k$-th person doesn't support a candidate, for $k = 1,\ldots, N$. Hence $\sum_{k=1}^NY_k$  stands for an unknown number of people in the population who support a candidate. Let us denote this number as $M$. The aim of the study is to estimate an unknown proportion (fraction) $\theta=  M/N$. A sample of size $n$ is drawn using simple random sampling without replacement scheme. In the sample number of people who support a candidate is observed and this number is a random variable. Let $\xi$ denote this random variable. The random variable $\xi$ has hypergeometric distribution (Barnett, 1974; Greene and Wellner, 2017) and its probability distribution function is
$$P_{\theta,N,n}\left\{\xi=x\right\}=\frac{\binom{\theta N}{x}\binom{(1-\theta)N}{n-x}}{\binom{N}{n}},\eqno{(1)}$$
for integer $x$ from set $\left\{\max\{0, n-(1-\theta)N\},\ldots, \min\{n, \theta N\}\right\}$. Unbiased estimator with minimal variance of the parameter $\theta$ is $\hat\theta=\frac{\xi}{n}$ (Cochran, 1977; Steczkowski, 1995). Variance of that estimator equals $D_\theta^2=\frac{1}{n^2}D_\theta^2\xi=\frac{\theta(1-\theta)}{n}\frac{N-n}{N-1}$ for all $\theta$. It is obvious the worst variance occurs when $\theta$ equals $1/2$.

\bigskip
{\bf 2. Stratified estimator.}
In some cases, the population of the study is strongly variable and support for a particular candidate depends on e.g. region or gender of voters. Therefore the sample is drawn due to simple random sampling without replacement scheme, so it can contain only people who support a candidate. To avoid this, stratified random sampling is used. This method of sampling assumes a division of the population among disjoint strata. After such division of the population, random sample is taken in each strata (Cochran, 1977). Let us divide the population $\cal U$ into two disjoint strata ${\cal U}_1$ and ${\cal U}_2$, ${\cal U}={\cal U}_1\cup{\cal U}_2$ of $N_1$ and $N_2$ people, respectively. The details of division of the population in stratification can be found in (Horgan, 2006; Hidiroglou and Kozak, 2017). For example, support in elections can depend on dominant political option at the time. In each strata fraction of people who support a candidate equals $\theta_1$ and $\theta_2$, respectively. The aim of the study is still to estimate the overall proportion $\theta$, not $\theta_1$ and $\theta_2$. Let $w_1$ denote a contribution of the first strata, i.e. $w_1=N_1/N$. Obviously the overall proportion equals
$$\theta=w_1\theta_1+w_2\theta_2,\eqno{(2)}$$
where $w_2=1-w_1$ is a contribution of the second strata. It seems intuitively obvious to take as our estimate of $\theta$,
$$\hat\theta_w =w_1\frac{\xi_1}{n_1} +w_2\frac{\xi_2}{n_2} ,\eqno{(3)}$$
where $n_1$ and $n_2$ denote sample sizes from the first and the second strata, respectively. Now, there are two random variables describing the number of units with a particular attribute in samples drawn from each strata:
$$\xi_1\sim H(N_1,\theta_1 N_1,n_1),\ \xi_2\sim H(N_2,\theta_2 N_2,n_2 ). \eqno{(4)}$$
Let us consider costs of sampling. Suppose that cost of sampling from the first strata equals $c_1$ and from the second one $c_2$. Funds for the poll are limited. Cost function is of the form:
$$C=c_1 n_1+c_2 n_2.	\eqno{(5)}$$
The main goal is to estimate the overall fraction $\theta$, not fraction in each strata. The parameter $\theta_1$ will be considered as a nuisance one. This parameter will be eliminated by appropriate averaging. Note that for a given $\theta\in[0,1]$, parameter $\theta_1$ is a fraction $M_1/N_1$ (it is treated as the number, not as the random variable) from the set
$$A=\left\{a_\theta,a_\theta+ \frac{1}{N_1},\ldots,b_\theta\right\}, \eqno{(6)}$$
where
$$a_\theta=\max\left\{0,\frac{\theta-w_2}{w_1}\right\}\hbox{ and } b_\theta=\min\left\{1,\frac{\theta}{w_1}\right\}\eqno{(7)}$$
and let $L_\theta$ be cardinality of $A$.

It is facile to prove that estimator $\hat\theta_w$ is an unbiased estimator of fraction $\theta$ (Sieradzki and Zieli\'nski, 2017). Hence it is necessary to compare variances of both estimators. Averaged variance of estimator $\hat\theta_w$ having regard to cost, could be as follows:
$$\eqalign{
D_\theta^2\hat\theta_w=&\frac{1}{L_\theta}\sum_{\theta_1\in A}\left(\frac{w_1^2}{n_1}\theta_1(1-\theta_1)\frac{N_1-n_1}{N_1-1}\right.\cr
&\left.+\frac{w_2^2}{(C-c_1 n_1)/c_2}\frac{\theta-w_1\theta_1}{w_2}
 \left(1-\frac{\theta-w_1\theta_1}{w_2}\right)\frac{N_2-(C-c_1 n_1)/c_2}{N_2-1}\right)\cr
 }\eqno{(8)}$$
Detailed analysis of variance $D_\theta^2\hat\theta_w$ can be found in (Sieradzki and Zieli\'nski, 2017; Sieradzki and Zieli\'nski, 2018). In further steps: firstly, finding 'the worst' situation, i.e. such value of proportion for which variance $D_\theta^2\hat\theta_w$ takes on its maximal value is needed. Then it is necessary to find such $(n_1^{opt},n_2^{opt})$, that minimizes this maximal variance. The optimal allocation of the sample is $(n_2^{opt}=(C-c_1 n_1^{opt})/c_2)$:
$$n_1^{opt}=\cases{
\frac{C\sqrt{N_2-1}w_1}{c_1\sqrt{N_2-1}w_1-\sqrt{c_1 c_2 w_2 (N(w_1^2-3w_1+1.5)-w_1)}},& for $w_1<w_1^*$,\cr
\hbox{numerical solution available},&for $w_1>w_1^*$,\cr}
\eqno{(9)}$$
where $w_1^*$ equals about $0.46$ (Sieradzki and Zieli\'nski, 2018).

In order to compare effectiveness of both estimators, it is necessary to determine sample size for the classical estimator $\hat\theta_c$. Let $n_c$ denote a sample size for estimator $\hat\theta_c$. Sample size could be described as follows (Sieradzki and Zieli\'nski, 2018):
$$n_c=\frac{C}{w_1 c_1+w_2 c_2}.	\eqno{(10)}$$
Example of application of this sample allocation will be considered in the next section.

\bigskip
{\bf 3. Example.}
Suppose that the aim of the research is to estimate support for a candidate (it will be referred to as a candidate "A") in second round of presidential elections in Poland. In Poland there are more than $30$ milion people who are entitled to vote (due to official statistics, in $2015$ there were $N=30709281$ voters). The standard way of estimation $\theta$ is to take a sample of size $n_c$ due to the scheme of simple sampling without replacement. In the sample the number of answers "yes, I will vote for candidate A" is counted. Let us denote this number as $\xi$. Obviously the standard estimator of the support is $\xi/n_c$.

In $2015$ some party which is linked with candidate 'A' won in $7$ of $16$ voivodeships. In those voivodeships there were $14526524$ people who may vote. In the remaining ones there were $16182757$ voters. Hence let us divide the population of electorate into two strata: the first one of the weight $w_1=14526524/30709281=0.47$ and the second one of the weight $w_2=16182757/30709281=0.53$. Suppose that costs of sampling from the first and the second strata equal $c_1=3$ and $c_2=1$, respectively. Funds for the sampling for this poll equal e.g. $C=1200$. These are exemplary values of these magnitudes, but for all values sample allocation is calculated in the same way. Sample size $n_c$ equals $618$. The optimal division $(n_1^{opt},n_2^{opt})$ of the sample for this numerical case could be calculated. After some calculations (which can be done in e.g. Mathematica) optimal sample allocation is obtained: $n_1=242$ and $n_2=474$.

Suppose that in the whole sample $100$ 'yes' answers were obtained. The point estimate of the support with classical estimator equals $\hat\theta_c=100/618 = 16.18\%$ and its estimated variance equals
$${\hat v}_c (100)=\frac{\hat\theta_c (1-\hat\theta_c)}{618}  \frac{30709281-618}{30709281-1}=0.00021946,	\eqno{(11)}$$
Suppose that in the sample of size $n_1$ from the first strata there were $10$ 'yes' answers and the number of 'yes' answers in the sample of size $n_2$ equals $128$. The point estimate of the support would be $\hat\theta_w=16.14\%$. The estimated variance of the estimator $\hat\theta_w$ equals
$$\eqalign{
\hat v(10,128)=&\left(\frac{14 526 524}{30 709 281}\right)^2\frac{\frac{10}{242}\left(1-\frac{10}{242}\right)}{616} \frac{14 526 524-242}{14 526 524-1}\cr
&+\left(\frac{16 182 757}{30 709 281}\right)^2 \frac{\frac{128}{474}\left(1-\frac{128}{474}\right)}{474} \frac{16 182 757-474}{16 182 757-1}=0.00001516.\cr}
\eqno{(12)}$$
The relative reduction of estimated variance equals
$$reduction=\left(1- \frac{{\hat v}_w (10,128)}{{\hat v}_c(100)}\right)\cdot100\%=30.94\%. \eqno{(13)}$$
Table 1 shows other possible results of the poll, assuming that the overall 'yes' answers equal to $100$, total funds equal $1200$, costs of sampling from the first and the second stratum equal $3$ and $1$, respectively.
 \vskip3truecm
{\bf Table 1.} Possible results for $\xi=100$, ${\hat v}_c(100)=0.00021946$, $c_1=3$, $c_2=1$, $C=1200$, $n_1=242$, $n_2=474$, $n_c=618$, $\hat\theta_c=16.18\%$.
$$\vbox{\tabskip=1em\offinterlineskip\halign{
\strut\hfil$#$\hfil&&#\vrule&\hfil$#$\hfil\cr
\xi_1&&\xi_2&&support&&variance&&reduction\cr\noalign{\hrule}
10&&128&&16.14\%&&0.0001516	&&30.94\%\cr
20&&111&&16.22\%&&0.0001750	&&20.28\%\cr
30&&93&&16.19\%	&&0.0001930	&&12.08\%\cr
40&&75&&16.16\%	&&0.0002061	&&6.10\%\cr
50&&57&&16.13\%	&&0.0002143	&&2.33\%\cr
60&&40&&16.21\%	&&0.0002188	&&0.31\%\cr
70&&22&&16.18\%	&&0.0002174	&&0.94\%\cr
80&&4 1&&6.15\%	&&0.0002112	&&3.77\%\cr
}}$$
In Tables 2 and 3 there are given possible results assuming, that the overall positive answers is $300$ and $400$ respectively.

{\bf Table 2.} Possible results for $\xi=200$, ${\hat v}_c(200)=0.000354193$, $c_1=3$, $c_2=1$, $C=1200$, $n_1=242$, $n_2=474$, $n_c=618$, $\hat\theta_c=32.36\%$.
$$\vbox{\tabskip=1em\offinterlineskip\halign{
\strut\hfil$#$\hfil&&#\vrule&\hfil$#$\hfil\cr
\xi_1&&\xi_2&&support&&variance&&reduction\cr\noalign{\hrule}
10 &&274&&32.31\%	 && 0.0001788	 &&49.53  \%\cr
20 &&257&&32.39\%	 && 0.000215	 &&39.29  \%\cr
30 &&239&&32.36\%	 && 0.0002466	 &&30.37  \%\cr
40 &&221&&32.33\%	 && 0.0002733	 &&22.83  \%\cr
50 &&204&&32.41\%	 && 0.0002954	 &&16.6   \%\cr
60 &&186&&32.38\%	 && 0.0003125	 &&11.78  \%\cr
70 &&168&&32.35\%	 && 0.0003247	 &&8.32   \%\cr
80 &&150&&32.32\%	 && 0.0003321	 &&6.24   \%\cr
90 &&133&&32.40\%	 && 0.0003352	 &&5.37   \%\cr
100&&115&& 32.37\%   &&	0.0003329	 &&6.01    \%\cr
110&& 97&& 32.33\%	 &&	0.0003258 	 &&8.02     \%\cr
120&& 80&& 32.41\%	 &&	0.0003146 	 &&11.17    \%\cr
130&& 62&& 32.38\%	 &&	0.0002979 	 &&15.89    \%\cr
140&& 44&& 32.35\%	 &&	0.0002763 	 &&21.99    \%\cr
150&& 26&& 32.32\%	 &&	0.0002498 	 &&29.46    \%\cr
160&& 9	&&32.40\%	 && 0.0002197	 &&37.97  \%\cr
}}$$
\vskip4.5truecm
{\bf Table 3.} Possible results for $\xi=300$, ${\hat v}_c(300)=0.000404188$, $c_1=3$, $c_2=1$, $C=1200$, $n_1=242$, $n_2=474$, $n_c=618$, $\hat\theta_c=48.54\%$.
$$\vbox{\tabskip=1em\offinterlineskip\halign{
\strut\hfil$#$\hfil&&#\vrule&\hfil$#$\hfil\cr
\xi_1&&\xi_2&&support&&variance&&reduction\cr\noalign{\hrule}
10&&421&&48.59\%&&0.0000947&&76.57\%\cr
20&&403&&48.56\%&&0.0001447&&64.19\%\cr
30&&385&&48.53\%&&0.0001899&&53.01\%\cr
40&&367&&48.5\%&&0.0002303&&43.03\%\cr
50&&350&&48.58\%&&0.0002652&&34.4\%\cr
60&&332&&48.55\%&&0.0002959&&26.8\%\cr
70&&314&&48.52\%&&0.0003217&&20.41\%\cr
80&&297&&48.6\%&&0.0003424&&15.29\%\cr
90&&279&&48.57\%&&0.0003586&&11.28\%\cr
100&&261&&48.54\%&&0.0003699&&8.47\%\cr
110&&243&&48.51\%&&0.0003764&&6.87\%\cr
120&&226&&48.59\%&&0.0003781&&6.45\%\cr
130&&208&&48.55\%&&0.000375&&7.22\%\cr
140&&190&&48.52\%&&0.000367&&9.2\%\cr
150&&172&&48.49\%&&0.0003541&&12.38\%\cr
160&&155&&48.57\%&&0.0003368&&16.66\%\cr
}}$$

Tables 4-6 contain proper columns, assuming that cost of sampling in the second strata is
greater than in the first strata, i.e. $c_1=1$ and $c_2=3$.

{\bf Table 4.} Possible results for $\xi=100$, ${\hat v}_c(100)=0.00021946$, $c_1=1$, $c_2=3$, $C=1200$, $n_1=405$, $n_2=265$, $n_c=582$, $\hat\theta_c=17.18\%$.
$$\vbox{\tabskip=1em\offinterlineskip\halign{
\strut\hfil$#$\hfil&&#\vrule&\hfil$#$\hfil\cr
\xi_1&&\xi_2&&support&&variance&&reduction\cr\noalign{\hrule}
10&&81&&17.22\%&&0.0002342&&4.23\%\cr
20&&75&&17.2\%&&0.0002372&&2.98\%\cr
30&&69&&17.19\%&&0.0002385&&2.45\%\cr
40&&63&&17.17\%&&0.0002381&&2.63\%\cr
50&&57&&17.16\%&&0.0002359&&3.52\%\cr
60&&51&&17.14\%&&0.000232&&5.13\%\cr
70&&45&&17.12\%&&0.0002263&&7.45\%\cr
80&&39&&17.11\%&&0.0002189&&10.49\%\cr
}}$$
\vskip4.3truecm
{\bf Table 5.} Possible results for $\xi=200$, ${\hat v}_c(200)=0.000387547$, $c_1=1$, $c_2=3$, $C=1200$, $n_1=405$, $n_2=265$, $n_c=582$, $\hat\theta_c=34.36\%$.
$$\vbox{\tabskip=1em\offinterlineskip\halign{
\strut\hfil$#$\hfil&&#\vrule&\hfil$#$\hfil\cr
\xi_1&&\xi_2&&support&&variance&&reduction\cr\noalign{\hrule}
10&&274&&32.31\%&&0.0001788&&49.53\%\cr
20&&257&&32.39\%&&0.000215&&39.29\%\cr
30&&239&&32.36\%&&0.0002466&&30.37\%\cr
40&&221&&32.33\%&&0.0002733&&22.83\%\cr
50&&204&&32.41\%&&0.0002954&&16.6\%\cr
60&&186&&32.38\%&&0.0003125&&11.78\%\cr
70&&168&&32.35\%&&0.0003247&&8.32\%\cr
80&&150&&32.32\%&&0.0003321&&6.24\%\cr
90&&133&&32.4\%&&0.0003352&&5.37\%\cr
100&&115&&32.37\%&&0.0003329&&6.01\%\cr
110&&97&&32.33\%&&0.0003258&&8.02\%\cr
120&&80&&32.41\%&&0.0003146&&11.17\%\cr
130&&62&&32.38\%&&0.0002979&&15.89\%\cr
140&&44&&32.35\%&&0.0002763&&21.99\%\cr
150&&26&&32.32\%&&0.0002498&&29.46\%\cr
160&&9&&32.4\%&&0.0002197&&37.97\%\cr
}}$$

{\bf Table 6.} Possible results for $\xi=300$, ${\hat v}_c(300)=0.000429142$, $c_1=1$, $c_2=3$, $C=1200$, $n_1=405$, $n_2=265$, $n_c=582$, $\hat\theta_c=51.55\%$.
$$\vbox{\tabskip=1em\offinterlineskip\halign{
\strut\hfil$#$\hfil&&#\vrule&\hfil$#$\hfil\cr
\xi_1&&\xi_2&&support&&variance&&reduction\cr\noalign{\hrule}
10&&254&&51.49\%&&0.0000548&&87.23\%\cr
20&&248&&51.48\%&&0.0000886&&79.36\%\cr
30&&242&&51.46\%&&0.0001206&&71.89\%\cr
40&&237&&51.64\%&&0.0001479&&65.54\%\cr
50&&231&&51.63\%&&0.0001766&&58.85\%\cr
60&&225&&51.61\%&&0.0002036&&52.56\%\cr
70&&219&&51.6\%&&0.0002289&&46.67\%\cr
80&&213&&51.58\%&&0.0002524&&41.2\%\cr
90&&207&&51.56\%&&0.0002741&&36.13\%\cr
100&&201&&51.55\%&&0.0002941&&31.46\%\cr
110&&195&&51.53\%&&0.0003124&&27.21\%\cr
120&&189&&51.52\%&&0.0003289&&23.36\%\cr
130&&183&&51.5\%&&0.0003437&&19.92\%\cr
140&&177&&51.49\%&&0.0003567&&16.88\%\cr
150&&171&&51.47\%&&0.000368&&14.25\%\cr
160&&165&&51.45\%&&0.0003775&&12.03\%\cr
}}$$

{\bf Summary.}
In the article an example of application of averaged sample allocation was presented. The classical estimator and stratified estimator were compared with respect to their estimated variances. The variance of estimator $\hat\theta_w$ depends strongly on costs of sampling $c_1$, $c_2$ and limited funds $C$. In the numerical study it was shown that whatever value $c_1$, $c_2$ and $C$ have, the estimated variance of $\hat\theta_w$ is smaller than estimated variance of $\hat\theta_c$. The reduction of the variance is up to $90\%$. It is worth noting that proposed sample allocation does not need any preliminary investigation, which is necessary in the case of Neyman allocation.

{\bf References}

Barnett, V. (1974). Elements of Sampling Theory, The English Universities Press Ltd.

Bartkowiak, R. (2003). Historia my\'sli ekonomicznej, Polskie Wydawnictwo Ekonomiczne, Warszawa.

Chaudhuri, A., Christofides, T., C., \& Saha, A. (2009). Protection of privacy in efficient application of randomized response techniques, Statistical Methods and Applications, 18, 389-418.

Cochran, W., G. (1977). Sampling Techniques (3rd ed.), New York: John Wiley.

Greene, E., \& Wellner, J., A. (2017). Exponential bounds for the hypergeometric distribution, Bernoulli, 23, 1911-1950.

Hada\'s-Dyduch, M. (2015). Polish macroeconomic indicators correlated-prediction with indicators of selected countries, In: Papie\.z, M. \& and \'Smiech, S. 9th Professor Aleksander Zelias International Conference on Modelling and Forecasting of Socio-Economic Phenomena. Conference Proceedings. Cracow: Foundation of the Cracow, 68-76.

Hansen, M., H., \& Hurwitz, W., N. (1946). The problem of non-response in sample surveys, Journal of the American Statistical Association, 41, 236, 517-529.

Hidiroglou, M., A., \& Kozak, M. (2017). Stratification of Skewed Populations: A Comparison of Optimisation-based versus Approximate Methods, International Statistical Review, 86, 87-105.

Roszkowska-M{\c a}dra, B., \& Ma\'nkowski D., R. (2010). Determinanty decyzji rolnik\'ow o korzystaniu z funduszy unii europejskiej i kredyt\'ow na dzia³alno\'s\'c rolnicz¹: przyk{\l}ad dla rolnictwa z rozwiniêtym systemem produkcji mlecznej w wojew\'odztwie podlaskim, Roczniki Nauk Rolniczych. Seria G, Ekonomika Rolnictwa, 97, 14-27.

Sieradzki, D., \& Stefa\'nczyk J. (2017). The conversion of the area of ecological crops in the selected EU states. Economic Science for Rural Development, 44, 190-196.

Sieradzki, D., \& Zieli\'nski W. (2017). Sample allocation in estimation of proportion in a~finite population divided into two strata. Statistics in Transition new series, 18(3), 541-548, 10.21307.

Sieradzki, D., \& Zieli\'nski W. (2018). Cost Issue in Estimation of Proportion in a Finite Population Divided Among Two Strata. submitted.

Steczkowski, J. (1995). Metoda reprezentacyjna w badaniach zjawisk ekonomiczno-spo{\l}ecz\-nych. PWN, Warszawa.

Szreder M. (2010). Metody i techniki sonda{\.z}owych bada\'n opinii. Wydawnictwo UEK, Kraków.

\bye